\documentstyle[tighten,preprint,eqsecnum,aps,prd]{revtex}

\def\be{\begin{equation}}
\def\ee{\end{equation}}

\begin{document}
\draft

\preprint{\vbox{\baselineskip=12pt
\rightline{IUCAA-26/97}
\rightline{}
\rightline{hep-th/9704175}
}}
\title{Solving the Graceful Exit Problem in Superstring Cosmology
\footnote{Based on talk given at Conference on
Big Bang and Alternative Cosmologies: A Critical Appraisal, Jawaharlal Nehru
Centre for Advanced Scientific Research, Bangalore, India, January 1997. To 
appear in the proceedings to be published by the Journal of Astrophysics
and Astronomy (The Indian Academy of Sciences).}}
\author{Sukanta Bose\footnote{Electronic address:
{\em sbose@iucaa.ernet.in}}}
\address{Inter-University Centre for Astronomy and Astrophysics, Post Bag 4, 
Ganeshkhind,\\ Pune 411007, India}

\date{March 1997}

\maketitle
\begin{abstract}

We briefly review the status of the ``graceful exit'' problem in superstring 
cosmology and present a possible resolution. It is shown that there exists
a solution to this problem in two-dimensional dilaton gravity provided 
quantum corrections are incorporated. This is similar to the recently proposed
solution of Rey. However, unlike in his case, in our one-loop corrected model
the graceful exit problem is solved for any finite number of massless scalar 
matter fields present in the theory.

\end{abstract}
\pacs{Pacs: 98.80.Cq, 04.60.Kz, 11.25.-w}

\narrowtext

\vfil
\pagebreak

\section{Introduction}

The Standard Cosmological Model successfully explains many features
related to the observed universe. However, it
does not offer a solution to the initial singularity problem or account
for the homogeneity and isotropy of the universe unless one invokes an
ad hoc inflaton field and fine-tunes the initial conditions.
Superstring cosmology appears to be promising in this regard. First, it
is known to be well behaved at ultraviolet energy scales or atmost have
mild singularities (Jain 1997). Second, apart from the graviton, it has 
a naturally occurring scalar field, the dilaton, whose kinetic 
energy can be used to 
drive the universe through an inflationary phase (Veneziano, 1996) . 
The solutions to the tree level effective action depict an FRW phase as 
well. Unfortunately, the tree level solution does not describe a smooth
singularity-free transition from the inflationary phase to the FRW phase.
This is called the ``graceful exit'' problem in superstring cosmology.

A quantum cosmology approach does indicate the possibility of a graceful 
exit (Gasperini et al 1996; Maharana et al 1997). More significantly, it
was shown by S.-J. Rey (1996) that this problem is avoided in a
string-inspired two-dimensional cosmological model provided back reaction
effects to first order in the Planck constant are incorporated. However,
Rey's solution unphysically requires that the number of
massless scalar fields $N$ in the universe be less than twenty four! Some
attempts, notably by Gasperini and Veneziano (1996), did not succeed in 
solving this problem. 

In this paper, we begin by briefly describing the salient features of 
four-dimensional superstring cosmology in section \ref{sec:qst}. We present a 
string-inspired classical action in two-dimensional dilaton gravity 
and illustrate the graceful exit problem in the context of its
cosmological solutions in section \ref{sec:CGHS}. In section \ref{sec:1loop}
we study the back-reaction effects due to one-loop quantum corrections
on the spacetime geometry. With the addition of our choice of one-loop 
counterterms to
the action we solve the graceful exit problem for any finite positive
value of N. We conclude this paper with a discussion on the implications
of our solution and possible scope for future research.

\section{Superstring Cosmology}
\label{sec:qst}

The low energy limit of string theory is given 
by an effective action of the type
\begin{equation}
\label{S4D}
S_{\rm eff} = {1\over 2} \int d^4 x \> \sqrt{-g} \> e^{-2\phi} \>
  \left[ \lambda_s^{-2} ({\cal R} + 4 \partial_\mu \phi \partial^\mu \phi )
 -{1\over 12} H_{\mu \nu \lambda} H^{\mu \nu \lambda} + V \right]
\ \ ,
\end{equation}
where $\phi$ is the dilaton field, ${\cal R}$ is the four-dimensional Ricci
scalar, $H_{\mu \nu \lambda}$ is the third-rank antisymmetric tensor field,
and $V$ is a term in which a dilaton potential or a cosmological constant 
term can be absorbed. There are two expansion parameters in the above action:
$e^{2\phi}$ is the analogue of the Planck's constant in quantum field theory 
and governs higher genus corrections. On the other hand $\lambda_s^2$ is 
related to the inverse of the string tension and controls string-size effects.
This action resembles the Brans-Dicke action with $\omega = -1$. Just as
in Brans-Dicke, here too calculations can be done in either the string 
(i.e., the Brans-Dicke) frame or the Einstein frame. The metric that appears
in the above action is the string metric and in this paper we will base our 
discussion in this frame.

The above action has been shown to have cosmological solutions (for reviews 
see Veneziano (1996) and Gasperini (1996)). Typically a solution exhibits
two branches defined by the range of the cosmic time $\tau$. The branch
corresponding to the range $-\infty < \tau \leq 0$ describes a 
superinflationary phase,
where the scale factor grows as an inverse power-law in cosmic time. This
phase is characterised by accelerated expansion and growing curvature. The
branch with $0 \leq \tau < \infty$ describes an FRW phase.

A crucial problem facing string cosmology is the lack of a smooth
transition from the superinflationary phase to the FRW phase. This is because
the superinflationary phase ends up in a region of diverging scalar curvature 
and coupling. This is the graceful exit problem in
the context of superstring cosmology. There are no-go theorems (Brustein \&
Veneziano 1994; Kaloper et al 1995) that show that even in the 
presence of realistic dilaton 
potentials a graceful exit from accelerated inflation does not occur, without 
invoking corrections from string-size effects (see, however, Kalyana Rama 
1997). In the subsequent sections we address this issue in cosmological 
models of string-inspired two-dimensional dilaton gravity.

\section{Classical Two-dimensional Cosmology}
\label{sec:CGHS}

A classical two-dimensional (2D) theory that describes cosmological models
of interest is given by the action of Callan et al (1992):
\begin{equation}
\label{S0}
S_0 = {1\over 2\pi} \int d^2 x \>\sqrt{-g} \left\{ e^{-2\phi} [R +
      4 (\nabla \phi)^2 - 4\Lambda ] - {1\over 2} \sum_{i=1}^N 
(\nabla f_i)^2 \right\}
\ \ ,
\end{equation}
where $\Lambda$ is a cosmological constant term, and 
$f_i$'s are $N$ massless scalar matter fields. In particular, the above action
has been shown to yield cosmological solutions with a superinflationary branch
disconnected from an FRW branch (Rey 1996; also see Mazzitelli \& Russo 1993).

Varying the above action with respect to the metric, dilaton, and the scalar
fields gives the following equations of motion:
\begin{eqnarray}
\label{classeq}
2 e^{-2 \phi} \left[ \nabla_\mu \nabla_\nu \phi + g_{\mu \nu}
( (\nabla \phi)^2 - \nabla^2 \phi + \Lambda ) \right] 
&+& {1\over 4} g_{\mu\nu} \sum_{i=1}^N (\nabla f_i)^2 - {1\over 2} 
\sum_{i=1}^N \nabla_\mu f_i \nabla_\nu f_i = 0 \>, \nonumber \\
e^{-2 \phi} \left[
R - 4 \Lambda + 4 \nabla^2 \phi - 4 (\nabla \phi)^2  \right] &=& 0 \>,
\quad {\rm and}  \quad \nabla^2 f_i = 0
\,.
\end{eqnarray}
In the conformal gauge, $g_{\mu \nu} \equiv e^{2\rho}\eta_{\mu \nu}$,
the metric components in the double null-coordinates, $x^\pm = t \pm x$, are
$g_{+-} = -{1 \over 2} e^{2 \rho}$ and $g_{++} = g_{--} = 0$.
In this gauge, the two-dimensional Ricci scalar is 
$R=8 e^{-2 \rho} \partial_+ \partial_- \rho$, and the
equations of motion take the form
\begin{eqnarray}
\label{classeqconf1}
\phi &:& \qquad  e^{-2(\phi+\rho)} \left[ -4 \partial_+ \partial_- \phi +
 4 \partial_+ \phi \partial_- \phi + 2
        \partial_+ \partial_- \rho - \Lambda e^{2 \rho} \right] = 0 \>, \\
\label{classeqconf2}
\rho &:& \qquad  e^{-2 \phi} \left[ 2 \partial_+ \partial_- \phi - 4 
\partial_+ \phi\partial_- \phi + \Lambda e^{2 \rho} \right]  =0
\,. 
\end{eqnarray}
Since we have gauge fixed
$g_{++}$ and $g_{--}$ to zero we must also impose their
equations of motion as constraints. This gives
\begin{equation}
\label{classconstraints}
e^{-2 \phi} ( 4 \partial_{\pm} \rho \partial_{\pm} \phi - 2 
{\partial_{\pm}}^2 \phi ) = -{1\over 2} \sum_{i=1}^N
\partial_{\pm} f_i \partial_{\pm} f_i
\,. 
\end{equation}
Adding Eqs. (\ref{classeqconf1}) and 
(\ref{classeqconf2}) gives the continuity equation
$\partial_+ \partial_- (\rho -\phi) = 0$, which shows
that $(\rho - \phi)$ behaves as a free field. 

In this paper we will discuss only the case of a vanishing cosmological 
term ($\Lambda = 0$). The homogeneous cosmological solutions are best 
described in terms of the new fields
$\Phi \equiv e^{-2\phi}$ and $\Sigma \equiv (\phi - \rho )$.
In vacuum, i.e., for $\dot{f}_i = 0$ (where an overdot denotes $\partial
/ \partial t$), for all values of $i$, the equations of
motion (\ref{classeqconf1}) simplify to $\ddot{\Phi} = \ddot{\Sigma} = 0$.
The corresponding solution is
\begin{equation}
- \Sigma \equiv (\rho - \phi ) = Q_{\Sigma } t + A \>, \qquad
\Phi \equiv e^{-2\phi} = Q_{\Phi} t + B \>,
\label{classsol}
\end{equation}
where $Q_{\Sigma }$, $Q_{\Phi}$, $A$, and $B$ are integration constants 
determined by initial conditions. On the other hand, the classical
constraints (\ref{classconstraints}) yield
\begin{equation}
\label{classcons}
\ddot{\Phi} + 2\dot{\Phi} \dot{\Sigma } = - Q_{\Phi} Q_{\Sigma } = 0
\ \ ,
\end{equation}
which shows that the cosmological solutions have two distinct branches
depending on whether $Q_{\Sigma }$ or $Q_{\Phi}$ is non-zero.

The solution in the first branch ($Q_{\Phi} \ne 0, Q_\Sigma = 0$) is given by
\begin{equation}
\label{inf}
\rho = \phi + \ln 2C ; \hskip1cm e^{-2\phi} =  - {8 C^2 t \over M} 
\ \ ,
\end{equation}
where $C$ and $M$ are constants. For real values of the coupling 
$e^{\phi}$ this solution describes the universe for only $t < 0$. From above, 
the spacetime metric for $t < 0$ is 
\begin{equation}
(ds)^2 = - \left( M \over -2 t \right) [dt^2 - dx^2] 
       = - [d \tau^2 - \left({M \over - \tau}\right)^2 dx^2] \>, 
\label{infmetric} 
\end{equation}
where $\tau \equiv - \left( - 2 M t \right)^{1/2}$ is the comoving time.
In the comoving coordinates the scale factor is $a(\tau)= M 
/(-\tau)$, which depicts a superinflationary evolution.

The solution in the second branch ($Q_\Sigma \ne 0, Q_\Phi = 0$) is given by
\begin{equation}
\label{milne}
\rho = \phi + \tilde{M} t ; \hskip1cm e^{-2\phi} = {\tilde C}^{-2}
\ \ ,
\end{equation}
which describes an ``expanding'' universe with the metric
\begin{equation}
\label{milnemetric}
(ds)^2 = - {\tilde C}^2 e^{2{\tilde M}t} [dt^2 - dx^2] = 
- [ d\tau^2 - ({\tilde M} \tau)^2 dx^2]
\ \ ,
\end{equation}
for non-negative values of the comoving time, $\tau \equiv 
({\tilde C}/ {\tilde M})\exp{\tilde M} t$.
The dilaton, and therefore the coupling, is constant in this branch of the
universe.

The two-dimensional universe described by Eqs. (\ref{infmetric}) and 
(\ref{milnemetric}) have two notable features. First, as in higher dimensions,
the two branches are related by scale factor duality (Veneziano 1991; Meissner
and Veneziano 1991; Sen 1991; Sen 1992; Hassan \& Sen 1991). Second, the 
superinflationary branch described by Eq. (\ref{infmetric})
terminates in a region of diverging scalar curvature and coupling as $\tau 
\to 0^-$. Thus, as in higher dimensions, a smooth transition from the 
superinflationary branch to the FRW phase does not occur in this classical 
theory. 

\section{Incorporating One-loop Corrections}
\label{sec:1loop}

In the superinflationary branch (\ref{infmetric}), well before  
$\tau \to 0^- $, the universe enters a region of strong 
coupling where corrections due to quantum gravitational effects become 
non-negligible. In this regime the predictions of the classical theory cannot 
be trusted and higher order corrections to the metric and the dilaton must 
be incorporated. Such an attempt was made by Mazzitelli \& Russo (1993)
and Rey (1996) by including one-loop corrections to the classical action.
The particular one-loop corrected model
they considered is the RST model (see Russo et al 1992).
Mazzitelli \& Russo showed that in the one-loop corrected model a 
smooth transition from the superinflationary phase to the FRW phase is not 
possible for negative values of $\Lambda $. However, Rey showed that  
the graceful exit problem is solved in this model for $\Lambda = 0$ {\em 
provided} the number of massless scalar fields $N$ is less than 24.

In this paper we propose the following one-loop corrected model, which is 
different from RST, and study its cosmological solutions:
\begin{equation}
S_1 = S_0 + {N\hbar \over 24\pi} \int d^2 x  \sqrt{- g} \>
( - {1\over 4} R \Box^{-1} R + 2(\nabla \phi)^2 - 3 \phi R )
\ \ ,
\label{1loopac}
\end{equation}
where $\Box_x G(x, x') =  \delta^2 (x-x') / 
\sqrt{-g(x)}$. The first term in the parenthesis is the Polyakov-Liouville
term that reproduces the trace anomaly for massless scalar fields (Callan et 
al 1992). However the one-loop action is defined only up to the addition of 
local covariant counterterms (Russo \& Tseytlin 1992). Our action differs
from RST only in the addition of different counterterms.
The higher order corrections beyond one loop are dropped by using the large 
$N$ approximation where $N \to \infty$ as $\hbar \to 0$ such that $\kappa 
\equiv N\hbar / 12$ remains finite. 

We now use the following one-loop corrected redefined fields
\begin{equation}
\label{qPS}
\Sigma \equiv (\phi - \rho) , \hskip1cm 
\Phi \equiv e^{-2 \phi} - \kappa\phi + {\kappa \over 2} \rho
       = e^{-2 \phi} - {\kappa \over 2} \rho -\kappa\Sigma  
\ \ .
\end{equation}
For homogeneous cosmologies with constant $f_i$'s, the equations of motion 
in terms of these 
variables take the same form as in the classical case, i.e., $\ddot{\Phi} =0
= \ddot{\Sigma}$. However the constraints get modified to 
\begin{equation}
\partial_{\pm}^2 {\Phi} + 2\partial_{\pm} {\Phi}\partial_{\pm} \Sigma 
= {3\over 2}\kappa \left[ \partial_{\pm}^2
\phi - 2\partial_{\pm} {\rho} \partial_{\pm} {\phi} \right] 
+\kappa t_{\pm} (x^{\pm}) 
\ \ ,
\label{qcons}
\end{equation}
where $t_{\pm} (x^{\pm})$ are nonlocal functions that arise 
from the homogeneous part of the Green function (see Callan et al 
1992; Bose et al 1995). The choice of these nonlocal 
functions determine the quantum state of the matter fields in the spacetime.
The total matter stress tensor can be expanded in orders of $\hbar$ as
$T^f_{\mu \nu} = (T^f_{\mu \nu})_{c\ell} + \langle T_{\mu \nu} \rangle$,
where $(T^f_{\pm \pm})_{c\ell} \equiv {1\over 2}\sum_{i=1}^N (\partial_{\pm} 
f_i )^2 $ and $\langle T_{\pm \pm} \rangle = \kappa [ \partial^2_{\pm} \rho -
(\partial_{\pm} \rho)^2 - t_{\pm}(x^{\pm}) ] $ is the one-loop contribution 
(Davies et al 1976). We will choose the state of the matter fields to be 
defined by
\begin{equation}
\label{qt}
t_{\pm} (x^{\pm}) = - {3\over 2} \left[ \partial_{\pm}^2
\phi - 2\partial_{\pm} {\rho} \partial_{\pm} {\phi} \right] \ \ .
\end{equation}

The equations of motion, $\ddot{\Phi} =0
= \ddot{\Sigma}$, yield the following solution
\begin{equation}
-\Sigma \equiv (\rho - \phi ) = Q_\Sigma t + A \>, \quad
\Phi \equiv e^{-2\phi} -{\kappa \over 2}\rho -\kappa \Sigma = 
Q_\Phi t + B \>. \label{qS} 
\end{equation}
Whereas the constraint (\ref{qcons}), under the condition (\ref{qt}), yields
the same classical expression (\ref{classcons}). Once again depending
on whether $Q_\Phi$ or $Q_\Sigma$ is non-zero, one finds two branches of
the solution. However, unlike the classical case, in this one-loop corrected
model each branch separately describes smooth transition from a 
superinflationary phase to an FRW phase.

The first branch is given by $\rho = \phi + \ln 2C$ and $e^{-2\phi}
-\kappa\rho / 2 = -8C^2 t/M$, where, unlike in Rey (1996), $\kappa$ is now 
positive. This solution can be reexpressed as 
\begin{equation}
e^{-2 \rho} - {\kappa \over 8 C^2} \rho = 
- {2  t \over M } \quad {\rm and} \quad
e^{-2 \phi} - {\kappa\over 2} \phi = - {8 C^2 t\over M} + 
{\kappa\over 2} \ln 2C \>.
\label{q1}
\end{equation}
This solution is valid for all real values of the conformal time $t$ and for
$\kappa > 0$.
At asymptotic past timelike infinity, $t \to -\infty$, the metric and the 
dilaton approach the forms (Rey 1996)
\begin{equation}
(ds)^2 \rightarrow -\left({M \over -2 t}\right) 
[dt^2 - dx^2] = - [d \tau^2 - \left({M \over -\tau} \right)^2 dx^2],
\quad {\rm and} \quad \phi \rightarrow -\ln (-2 \tau)
\ \ ,
\end{equation}
where $\tau \equiv -(-2Mt)^{1/2}$. As $t \to \infty$
\begin{equation}
(ds)^2 \rightarrow  - e^{32 C^2 t / \kappa M} 
[dt^2 - dx^2] = - [d \tau^2 - \left({16 C^2 \over \kappa M} \tau 
\right)^2 dx^2] \quad {\rm and} \quad \phi \rightarrow  \ln \tau 
\ \ ,
\end{equation}
where the comoving time is $\tau \approx (\kappa M/ 16 C^2) 
\exp (16 C^2 t / \kappa M)$. 
Thus in the one-loop corrected model the universe begins in a classical
superinflationary phase and ends up being in an FRW phase. 
The solution corresponding to the second branch is different from Rey's
and is discussed elsewhere (Bose \& Kar 1997).

The remaining question we would like to address is whether our one-loop 
corrected model indeed displays a smooth transition between the two phases
in each branch. This can be checked by verifying that the scalar curvature 
remains finite at all times and the coupling remains small always, such that
the large $N$ approximation is not violated. These requirements can be 
shown to hold for the one-loop corrected solution (\ref{q1}) (on the same 
lines as Rey (1996), but now for any finite positive value of $N$). A
detailed discussion of the exact solutions to our model (\ref{1loopac})
is given in Bose \& Kar (1997). There we also address the question of
how scale factor duality affects graceful exit.

\section{Discussion}
\label{sec:disc}

Above we proposed a two-dimensional model in dilaton gravity that solves 
the graceful exit problem for any finite positive value of $N$. As shown by
Eq. (\ref{qt}) this solution requires the presence of a homogeneous 
distribution of massless scalar matter fields. Further, it can be shown that
the Weak Energy Condition (WEC) is violated in this solution (Bose \& Kar 
1997). In fact this 
agrees with the recent result of Brustein \& Madden (1997) and Kar (1996).

The next logical step is to account for string-size corrections in the 
4D cosmological models. One expects that this might solve the 
graceful exit problem in 4D. However, a perturbative solution, as the one
given here for 2D, is unlikely to do the job. A more fruitful approach,
as advocated by Brustein and Veneziano (1994), might be to look for
a conformal field theory (possessing cosmological solutions)
endowed with appropriate duality symmetries that can be expolited 
to probe the strong coupling regions successfully. 

I am grateful to J. V. Narlikar and G. Veneziano for motivating me to look
into this problem. I thank S. Kalyana Rama, J. Maharana, T. Padmanabhan, 
S. Panda, V. Sahni, and especially S. Kar and S. Sinha for helpful 
discussions.

\end{document}